\documentclass[12pt]{article}
\usepackage{axodraw}

\def\ltsim{\lower3pt\hbox{$\, \buildrel < \over \sim \, $}}
\def\gtsim{\lower3pt\hbox{$\, \buildrel > \over \sim \, $}}
\def\be{\begin{equation}}
\def\ee{\end{equation}}
\def\ba{\begin{eqnarray}}
\def\ea{\end{eqnarray}}
\def\ga{\mathrel{\raise.3ex\hbox{$>$\kern-.75em\lower1ex\hbox{$\sim$}}}}
\def\la{\mathrel{\raise.3ex\hbox{$<$\kern-.75em\lower1ex\hbox{$\sim$}}}}

\begin{document}

\baselineskip=16pt 
\begin{titlepage}
\rightline{OUTP-01-16P}
\rightline{hep-th/0103184}
\rightline{March 2001}  
\begin{center}

\vspace{1.7cm}

\Large {\bf Bulk Fermions in Multi-Brane Worlds}

\vspace*{7mm}
\normalsize

{\large \bf {Stavros
Mouslopoulos\footnote{s.mouslopoulos@physics.ox.ac.uk} 
 }}

\vspace*{4mm}

{\large
\smallskip 
\medskip 
{\it Theoretical Physics, Department of Physics, Oxford University}\\
{\it 1 Keble Road, Oxford, OX1 3NP,  UK}
\smallskip}

\vskip0.5in \end{center}
 
\centerline{\large\bf Abstract}

We study bulk fermion fields in various multi-brane  models with localized
gravity. The chiral zero mode that these models support can be
identified as a right-handed sterile neutrino. In this case small
neutrino Dirac masses can naturally appear due
to a localization of the bulk fermion zero
mode wavefunction, in an analogous way to graviton, without
invoking the see-saw mechanism. The conditions and the options for
localization are discussed in detail. It is shown that,
considering a well motivated five dimensional mass term, the localization
behaviour of this mode can resemble
the graviton's at least in a region of the parameter space.
As a result, the $''+-+''$, $''++''$ models can support,
in addition to the ultralight graviton KK state, an
ultralight localized and strongly coupled  bulk fermion KK mode.
We find that there are severe constraints on the parameter space of
$''+-+''$ and $''++''$ models if the neutrino properties resulting
from this light fermion state are to be reasonable. 
Furthermore, in the case that also the Bigravity scenario is realized
the above special KK mode can induce too large mixing between the neutrino
and the KK tower sterile modes restricting even more the allowed  parameter space.

\vspace*{2mm} 

\end{titlepage}


\section{Introduction}

The study of bulk fermion  fields, although not something new \cite{Rubakov:1983bb}, turns out to be of particular interest 
in the context of brane-world scenarios both in the case of models
with large
extra dimensions \cite{Arkani-Hamed:1998rs,Arkani-Hamed:1999nn,Antoniadis:1998ig} (factorizable geometry) and in models of localized
gravity \cite{Gogberashvili:1998vx,Randall:1999ee,Randall:1999vf} (non factorizable geometry) since they can provide possible
new ways to explain the smallness of the neutrino masses, neutrino
oscillations and the pattern of
fermion mass hierarchy. 

In the context of string and M-theory, bulk fermions arise as
superpartners of gravitational moduli, such as, those setting the radii of
internal spaces. Given this origin, the existence of bulk fermions is
unavoidable in any supersymmetric string compactification
and represents a quite generic feature of string
theory\footnote{However, note that brane-world models with non
factorizable geometry have not yet been shown to have string
realizations. For string realizations of models with large extra
dimensions see Ref.\cite{Antoniadis:1998ig}}. This constitutes the most likely origin of such particles
within a fundamental theory and, at the same time, provides the basis
to study brane-world neutrino physics.

In the traditional approach the small neutrino masses are a result of
the see-saw mechanism, in which a large right-handed Majorana mass
$M_{R}$ suppresses the eigenvalues of the neutrino mass matrix
leading to the light neutrino mass $m_{\nu}\sim \frac{m_{fermion}^{2}}{M_{R}}$. The neutrino
mixing explanations of the atmospheric and solar neutrino anomalies
require that $M_{R}$ to be a superheavy mass scale $> 10^{10}$ GeV.     

In the case of large extra dimensions \cite{Arkani-Hamed:1998rs,Arkani-Hamed:1999nn,Antoniadis:1998ig}, despite  the absence
of  a high scale like $M_{R}$ (since in such models the fundamental scale can be
as low as 1 TeV), small neutrino masses \cite{Dienes:1999sb,Arkani-Hamed:1998vp} (Dirac or Majorana) can arise
from an intrinsically higher-dimensional mechanism. The idea is that any fermionic state
that propagates in the bulk, being a Standard Model (SM) singlet can
be identified with
a sterile neutrino which through it's coupling to the SM
left-handed neutrino can generate small neutrino mass. In
the case of factorizable geometry, the smallness of the induced masses  is due to the fact that
the  coupling is suppressed by the large volume of
the internal bulk manifold.
In other words, the interaction probability between the bulk fermion
zero mode, the Higgs
and Lepton doublet fields (which are confined to a brane) is small
because of the large volume of bulk compared to the thin wall where the SM
states are confined, resulting  a highly
suppressed coupling. In the context of these models one can attempt to
explain the atmospheric and solar neutrino anomalies (see e.g. \cite{Dienes:1999sb,Dvali:1999cn,Barbieri:2000mg,Lukas:2000wn,Lukas:2000rg}).

\begin{figure}[t]
\begin{center}
\begin{picture}(300,200)(0,50)

\SetWidth{2}
\Line(10,50)(10,250)
\Line(290,50)(290,250)

\SetWidth{0.5}
\Line(150,50)(150,250)
\Line(10,150)(290,150)

\Text(-10,250)[c]{$''+''$}
\Text(310,250)[c]{$''+''$}
\Text(170,250)[c]{$''-''$}


{\SetColor{Green}
\Curve{(10,240)(50,192)(65,181)(80,173)(95,167)(110,162)(130,157)(150,155)}
\Curve{(150,155)(170,157)(190,162)(205,167)(220,173)(235,181)(250,192)(290,240)}
}


{\SetColor{Red}
\DashCurve{(10,240)(50,192)(65,181)(80,173)(95,167)(110,161)(130,155)(150,150)}{3}
\DashCurve{(150,150)(170,145)(190,139)(205,133)(220,127)(235,119)(250,108)(290,60)}{3}
}


{\SetColor{Brown}
\DashCurve{(10,145)(15,149)(20,150)(40,153)(65,158)(140,209)(150,210)}{1}
\DashCurve{(150,210)(160,209)(235,158)(260,153)(280,150)(285,149)(290,145)}{1}
}
\end{picture}
\end{center}

\caption{The right-handed fermion zero mode  (solid line), first (dashed line) and second
(dotted line) KK states wavefunctions in the symmetric $''+-+''$
model. The same pattern can occur
for the corresponding wavefunctions in the $''++''$ model. The
wavefunctions of the zero and the first KK mode are localized on the
positive tension branes. Their absolute value differ only in the
central region where they are both suppressed resulting to a very
light first KK state.}

\end{figure}
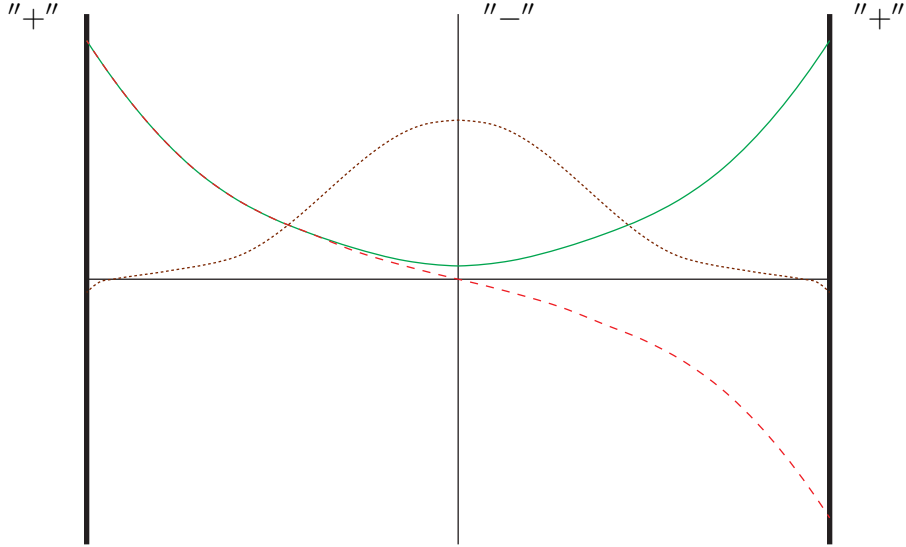

In the context of brane world models with localized gravity \cite{Gogberashvili:1998vx,Randall:1999ee,Randall:1999vf} (non
factorizable geometry) small neutrino masses can again be achieved,
without invoking the see-saw mechanism. In more detail, in this case  the
mechanism generating the small coupling between the Lepton doublet and
the Higgs which live on the brane and the  right-handed sterile neutrino zero mode is not due to the
compactification volume (which is now small)  but due to the fact that
the sterile neutrino wavefunction can be localized [16-28]
on a distant brane.
One may thus arrange that the overlap between this mode and the SM
brane is sufficiently small. 
In this case the $AdS_{5}$ geometry localizes the fermion zero mode on
negative tension branes.  
Localization can occur on positive branes if a mass term of
the appropriate form is added to counterbalance the effect of
the $AdS_{5}$ geometry, by applying the ideas presented
in Ref.\cite{Jackiw:1976fn,Rubakov:1983bb}. Such a mass term appears
naturally in the case that, the branes arise
as the limiting cases of domain walls that are created
from a five-dimensional scalar field with an nontrivial ground state
(kink or multi-kink) \cite{Kehagias:2000au,Kehagias:2000dg}. In this case the scalar field  naturally
couples to the bulk fermion through an non trivial ``mass'' term and
will naturally induce localization to the bulk modes on positive or
negative branes - which depends on the vacuum expectation value of the
scalar field.

In this paper we analyze the localization of a bulk fermion  the
mass spectrum and the coupling between SM neutrino and bulk states in
the context of multi-brane worlds (e.g. see Ref.\cite{Hatanaka:1999ac}). We discuss in detail the conditions
and the options for the localization in relation to the form of the
bulk mass term.
We also discuss  
the possibility of generating small neutrino masses in the context of
$''++-''$, $''+-+''$\footnote{We consider the $''+-+''$ configuration
as a toy-model ignoring the phenomenological difficulties associated
with the  presence of a negative tension brane \cite{Pilo:2000et} since it's
phenomenology is very similar to the $''++''$ which includes only
positive tension branes.}, $''++''$ models. We determine for which
regions of the parameter space lead to a solution of the
hierarchy problem and generation of small neutrino masses.
The study of $''+-+''$ and
$''++''$ models reveals the possibility of  an ultralight KK state of
the bulk fermion  analogous to the  KK graviton in the gravitational sector 
\cite{Kogan:2000wc,Mouslopoulos:2000er,Kogan:2000xc,Kogan:2000vb}.
This is due to the fact that, in this region the wavefunction of the
right-handed bulk fermion states obeys a similar equation to that of
the graviton. 
The  above fermion state, when exists, imposes even more severe
constraints on the parameter space of these
models. 

The organization of the paper is as follows: In the next section 
we review the general formalism, and discuss the form of the bulk
neutrino mass term which turns out to be
 critical in determining the neutrino properties. In section
3 we review the  bulk fermion properties in the Randall-Sundrum (RS) model \cite{Randall:1999ee}. In 
section 4 we extend this to  the case of three brane $''++-''$ model \cite{Kogan:2000xc}. In 
sections 5 and 6 we study in detail the ``Bigravity'' three brane
models $''+-+''$ \cite{Kogan:2000wc,Mouslopoulos:2000er}
and $''++''$ \cite{Kogan:2000vb}. In section 7 we  discuss the implications of the
realization of the ``Bigravity'' in relation to the presence of
bulk spinors. The overall implications and conclusions are presented
in section 8.


\section{General Framework}

Following the framework introduced in Ref.\cite{Grossman:2000ra}, we
consider a spinor $\Psi$  in a five dimensional $AdS_{5}$ space-time, where the
extra dimension is  compact and has the geometry of an orbifold
$S^{1}/Z_{2}$. The  $AdS_{5}$ 
background geometry is described by \footnote{We will assume that the
background metric is not modified by the presence of the bulk fermion,
that is, we will neglect the back-reaction on the metric from the bulk
fields.}:
\be
ds^2=e^{-2\sigma(y)}\eta_{\mu\nu}dx^\mu dx^\nu - dy^2
\ee
where the warp factor $\sigma(y)$  depends on the details of the model considered. For the moment we assume that
we have a model with a number of positive and negative tension flat
branes (the sum of the brane tensions should be zero if one wants 
flat four dimensional space on the branes) and that this function is known ( it can be found by
looking the system gravitationally). 

The action for a Dirac Spinor of a  mass $m$  in such a background is given
by \footnote{We do not include a Majorana mass term, ${\Psi^{T}}C\Psi$,
which is forbidden if the bulk fermion has a conserved lepton number.}:

\be
S=\int d^4 x \int dy \sqrt{G} 
\{  E^{A}_{\alpha}\left[
\frac{i}{2} \bar{\Psi} \gamma^{\alpha} \left( \overrightarrow{{\partial}_{A}}-\overleftarrow{\partial_{A}} \right)
\Psi + \frac{\omega_{bcA}}{8} \bar{\Psi} \{
\gamma^{\alpha},\sigma^{bc} \} 
\Psi \right] - m \Phi \bar{\Psi}\Psi \}
\ee
where 
$G=det(G_{AB})=e^{-8\sigma(y)}$.
The four dimensional representation of the Dirac matrices in
five-dimensional flat space is chosen to be:~$\gamma^{\alpha}=(\gamma^{\mu},i \gamma_{5})$.
The inverse vielbein is given by
$E^{A}_{\alpha}=diag(e^{\sigma(y)},e^{\sigma(y)},e^{\sigma(y)},e^{\sigma(y)},1)$.
Due to the fact that the vielbein is symmetric, the contraction of $\omega_{bcA}$
\footnote{where $\omega_{\mu ab}=\frac{1}{2}(\partial_{\mu}e_{b
\nu}-\partial_{\nu} e_{b \mu}){e_{a}}^{\nu}-\frac{1}{2}(\partial_{\mu}e_{a
\nu}-\partial_{\nu} e_{a \mu}){e_{b}}^{\nu}
-\frac{1}{2}{e_{a}}^{\rho}{e_{b}}^{\sigma}(\partial_{\rho}e_{c \sigma}-\partial_{\sigma}e_{c \rho}){e^{c}}_{\mu}$ } 
with the corresponding term in the action gives vanishing contribution.
The mass term is assumed to be generated by a Yukawa coupling with a
scalar field $\Phi$ which has a nontrivial stable vacuum $\Phi(y)$. 

A few comments on presence of the mass term are in order. 
The  motivation for introducing a mass term  of this form comes
from the need to  localize  fermion zero modes in the
extra dimension. This is discussed in more details in the next
paragraph.
Note that 
the ``localization'' of the wavefunction of a state does not
necessarily reflect the actual localization of the state, since in
one has to take in account the nontrivial geometry of the extra
dimension - something that is done when we calculate physical
quantities. Also when we note that a state is localized on a brane
we mean that this holds irrespectively if the space is compact or not
(thus the state should be normalizable even in the non-compact case). 
 In order to have a localized
state (zero mode) on a positive brane it is necessary to have an
appropriate bulk mass term.
This is because \cite{Bajc:2000mh,Chang:2000nh} the background $AdS_{5}$ geometry itself
has the opposite effect favouring localization on negative tension branes. 
As we will discuss, this leads to  a critical mass
$m_{cr}$ below which  the localization is still on the
negative brane (if we consider a configuration of a positive and one
negative brane), for $m=m_{cr}$ the is no localization and for
$m>m_{cr}$ the zero mode is localized on the positive tension brane
\footnote{One may ask why one should localize the fermion zero mode on
a positive tension brane and not on a negative ? In the case of RS
model it is obvious that if we demand to solve the hierarchy problem
and in the same time to create small neutrino masses through this
mechanism, one should localize the fermion zero mode on the positive
brane. In the case of
multi-brane models this is not a necessity and thus, in principle,
another  possibility (of course in this case the mass  term - if
needed - should have
different form (e.g. for $m \rightarrow -m$)).}. 

\subsection{The mass term}

Let us return to the specific form  of the mass term and it's generality
since this will become important in later discussions.
Since the mass term is a key element for the localization let us, for
a moment assume that it has the form $m\Phi(y) \bar{\Psi}\Psi$ where
$\Phi(y)$ is the vev of a scalar field which has a nontrivial stable vacuum (its
vev does not depend on  the remaining  spatial dimensions). 
Now if we do a simple calculation, in the case of flat extra dimension
(the general arguments will apply also in our background eq.(1), taking
in account also the effects of the $AdS_{5}$ geometry), we find that the above configuration
implies that the zero mode  satisfies a Schr\"ondiger equation with
potential of the form $V(y)=\Phi^{2}(y) -\Phi'(y)$. In
order to localize one state the profile of $\Phi(y)$ should be such
that it creates a  potential well. The way to do this is to assume that the
ground state of the scalar field has a kink or a multi-kink profile \cite{Jackiw:1976fn,Rubakov:1983bb,Kehagias:2000au}. Although
 the details of these profiles depend on the form of the
potential of the scalar field $\Phi$,  if we demand strong
localization of the states, the kink profiles tend to $\theta$-
functions. This implies that the function $\Phi$ in eq.(2) can be
considered  as an arbitrary combination of $\theta$-functions
(compatible with the symmetries of the action).   
However, as shown in Ref.\cite{Kehagias:2000au} the same field  $\Phi$
can be used in order to create the branes themselves. This restricts
the possible form of the mass term.  If we assume that
the same field $\Phi$ creates the branes and localizes the fermion
zero mode the mass term should have a (multi-)kink form, with
 $\Phi(y)=\frac{\sigma'(y)}{k}$  up to
a sign.\footnote{Note thought, that if we choose the
opposite sign for the mass term (i.e. $m \rightarrow -m$) the
localization of the fermion zero mode will always be on the negative
tension branes. Taking in account the fact that the latter occurs also,
in a region of the parameter space, in the case that our mass term
choice is the one that appears in eq.(2),  we will not consider this
possibility separately since we can easily, as we will see, generalize our results
for $m<0$ (the presence of such a mass term sharpens the localization
of the states on negative branes).}
 The previous argument also supports the $\theta$-function
form of the mass term (and not for example a $\tanh(y)$ profile) since we
assume that the branes are infinitely thin. Note thought that  in 
the $''++''$ model  due to the $AdS_{4}$
geometry on the branes the, $\sigma'(y)$ function does not have just a
$\theta$-function form  but it also involves  kink profiles.

 If one assumes that
the mass is generated by coupling to a scalar field, different from the
one that creates the branes, it can have any form allowed by the
dynamics and the symmetries of the action. For example it can take
 the form: $-m \left( \theta(y)-\theta(-y)) \right)
$. In this case, it  will tend to induce localization  on the brane
siting at the origin of the orbifold. Nevertheless  this will not be
satisfactory option in multi-brane models where the desired
$M_{Pl}/M_{EW}$ hierarchy is not
generated between the two first branes (e.g. $''+-+''$ or even in the
$''++''$ model)
since in these cases the will be not generally be possible to 
simultaneously solve the hierarchy problem and
generate small neutrino masses (it would work though in the $''++-''$
model - or it could work in the cases where the background induced
localization dominates but in that case the mass term would be
unnecessary anyway). In any case since we are interested in the most
economic, in terms of parameters and fields, models we will not
consider these possibilities.

The geometry we consider has a $Z_{2}$ symmetry ( $y \rightarrow -y$ ). 
Under this the fermion parity is defined as: $\Psi(-y)=\gamma_{5}\Psi(y)$
(i.e. $\Psi_{L}(-y)=-\Psi_{L}(y)$, $\Psi_{R}(-y)=\Psi_{R}(y)$ ) and
 changes the sign of a Lagrangian mass term of the form: $m\bar{\Psi}\Psi=m(\bar{\Psi}_{L}\Psi_{R}+\bar{\Psi}_{R}\Psi_{L})$.
The full mass term  however is invariant under the $Z_{2}$ since the function
$\sigma'(y)$ is also odd under the reflections $y \rightarrow -y$ . 
With this definition of
parity  one of the wavefunctions will be
symmetric and the other antisymmetric with respect to the center of the orbiford
. Note that this implies that the odd wavefunction will be zero
at the orbifold fixed points (i.e. zero coupling to fields confined to
that points).  
Since we would like in what follows to use the
right-handed component in order to give mass to SM neutrinos, which
could be  confined on a brane at an orbifold fixed point,  we choose the
right-handed wavefunction to be even (i.e. non-vanishing coupling)
and the left-handed to be odd.

\subsection{The KK decomposition}

It is convenient to write the action in terms of the fields: $\Psi_{R}$
and $\Psi_{L}$ where
$\Psi_{R,L}=\frac{1}{2}(1\pm \gamma_{5})\Psi$ and
$\Psi=\Psi_{R}+\Psi_{L}$.
The action becomes:

\ba
S=\int d^4 x \int dy  
\{  e^{-3\sigma}\left( \bar{\Psi}_{L}i \gamma^{\mu} \partial_{\mu} \Psi_{L} +
\bar{\Psi}_{R}i \gamma^{\mu} \partial_{\mu} \Psi_{R} \right) -  e^{-4\sigma} m
\left(\frac{\sigma'(y)}{k}\right) \left( \bar{\Psi}_{L}\Psi_{R} +
\bar{\Psi}_{R}\Psi_{L} \right) \nonumber \\ -\frac{1}{2}\left[ \bar{\Psi}_{L} (e^{-4\sigma}\partial_{y}+\partial_{y}e^{-4\sigma} ) \Psi_{R} -
\bar{\Psi}_{R}(e^{-4\sigma}\partial_{y}+\partial_{y}e^{-4\sigma} ) \Psi_{L}  \right] 
\ea
writing   $\Psi_{R}$ and
$\Psi_{L}$ in the form:
\be
\Psi_{R,L}(x,y)=\sum_{n}\psi^{R,L}_{n}(x)e^{2\sigma(y)}f_{n}^{R,L}(y)
\ee
the action can be brought in the form
\be
S=\sum_{n} \int d^4 x \{\bar{\psi}_{n}(x) i \gamma^{\mu} \partial_{\mu} \psi_{n}(x) -
m_{n}\bar{\psi}_{n}(x) \psi_{n}(x) \}
\ee
provided the wavefunctions obey the following equations
\ba
\left( -\partial_{y}
+m\frac{\sigma'(y)}{k}\right)f^{L}_{n}(y)=m_{n}e^{\sigma(y)}f^{R}_{n}(y)
\nonumber \\
\left( \partial_{y} +m\frac{\sigma'(y)}{k}\right)f^{R}_{n}(y)=m_{n}e^{\sigma(y)}f^{L}_{n}(y)
\ea
and  the orthogonality relations (taking  account of the $Z_{2}$ symmetry):
\be
\int_{-L}^{L} dy  e^{\sigma(y)} {f^{L}}^{*}_{m}(y) f^{L}_{n}(y)=\int_{-L}^{L} dy e^{\sigma(y)} {f^{R}}^{*}_{m}(y) f^{R}_{n}(y)=\delta_{mn}
\ee
where we assume that the length of the orbifold is $2L$.

We solve the above system of coupled differential equations by
substituting $ f^{L}_{n}(y)$ from the second in the first equation. Thus we end up
with a second order differential equation, which can always be brought
to a Schr\"ondiger form by a convenient coordinate transformation from y
to z coordinates  related through
$\frac{dz}{dy}=e^{\sigma(y)}$, the coordinate transformation  chosen to
 eliminate the terms involving first derivatives. 
Thus we end up with the differential equation of the form:

\be
\left\{-
\frac{1}{2}\partial_z^2+V_{R}(z)\right\}\hat{f}^{R}_{n}(z)=\frac{m_n^2}{2}\hat{f}^{R}
_{n}(z)
\ee

\be
{\rm with}\hspace*{0.5cm} V_{R}(z)=\frac{\nu(\nu+1)(\sigma'(y))^{2}}{2[g(z)]^2}-
\frac{\nu}{2[g(z)]^2}\sigma''(y)
\ee
Here $\hat{f}^{R}_{n}(z)=f^{R}_{n}(y)$ and we have defined $\nu\equiv\frac{m}{k}$ and
$g(z)\equiv e^{\sigma(y)}$. The left handed wavefunctions are given by
\footnote{Note that it can be shown that the left-handed component
obeys also a similar Schr\"odinger equation with $V_{L}(z)=\frac{\nu(\nu-1)k^{2}}{2[g(z)]^2}+
\frac{\nu}{2[g(z)]^2}\sigma''(y)$ which is the same as $V_{R}$ for
$\nu \rightarrow -\nu$. }:

\be
f^{L}_{n}(y)= \frac{e^{-\sigma(y)}}{m_{n}} \left( \partial_{y} +m\frac{\sigma'(y)}{k}\right)f^{R}_{n}(y)
\ee                   
The form of eq.(8)  and (9) are
 exactly the same as that satisfied by the graviton
when $\nu=\frac{3}{2}$ \cite{Kogan:2000wc}.
For any $\nu$, we note that before orbifolding  the system supports
two zero modes (left-handed and right-handed) \cite{Grossman:2000ra}. 
  However, the orbifold
compactification leaves only a chiral right-handed zero mode. 

We  note that since the bulk fermion mass, $m$, is a
parameter that appears in the original five dimensional Lagrangian
its ``natural'' value  is of the order of the five dimensional Planck
scale $M_{5}$. Now since we assume that  $k<M_{5}$ (in order to trust our
perturbative analysis when we consider the configuration
gravitationally) it is clear that the
``physical'' value of $\nu$ is $\nu>1$. However, we will always
comment on the behaviour of our results out of this region (even for
negative values). 
  
We are particularly interested in the coupling of the bulk spinor  to
the SM neutrinos since  this is the way that the neutrino masses will
be generated. In order to avoid weak scale neutrino masses and lepton
number violating interactions we assign lepton number $L=1$ to the
bulk fermion state and thus the only gauge invariant coupling is of
the form

\be
S_{Y}=-\int d^{4} x \sqrt{- g_{Br}}\{Y_{5}\bar{L_{0}}(x)\widetilde{H_{0}}(x)\Psi_{R}(x,L_{Br})+h.c.\} 
\ee
where $H$ is the SM Higgs field, $L$ is the SM lepton doublet, $\widetilde{H}=i\sigma_{2}H^{*}$, $g^{vis}_{\mu\nu}$ is the
induced metric on the brane and
$g_{vis}=det(g^{vis}_{\mu\nu})$. The Yukawa parameter $Y_{5}$ has mass dimension
$-\frac{1}{2}$ and thus since it appears a parameter in the five
dimensional action, its ``natural'' value is  $Y_{5} \sim
\frac{1}{\sqrt{M_{5}}} \sim \frac{1}{\sqrt{k}}$.

 To obtain canonical normalization for the kinetic terms of the SM
neutrino we perform the following field rescalings $H_{0}\rightarrow
e^{\sigma(L_{Br})} H$,~$L_{0}\rightarrow
e^{3 \sigma(L_{Br})/2} L$ , where $L_{Br}$ is the position of the
brane that SM is confined. This gives 

\be
S_{Y}=- \sum_{n\ge0}\int d^{4} x \{y_{n}\bar{L}(x)\widetilde{H}(x){\psi_{n}}^{R}(x)+h.c.\} 
\ee
where 
\be
y_{n} \equiv e^{\sigma(L_{Br})/2}~Y_{5}~f^{R}_{n}(L)=(g(z_{Br}))^{1/2}~Y_{5}~{\hat{f}^{R}}_{n}(z_{L})
\ee
From the above interaction terms we can read off the mass matrix
$\mathbf{M}$ that appears in the Lagrangian as  $\bar{\psi}^{\nu}_{L}\mathbf{M}\psi^{\nu}_{R}+ h.c.$
where we have defined
$\psi^{\nu}_{L}=(\nu_{L},\psi_{1}^{L},\ldots,\psi_{n}^{L})$
and
$\psi^{\nu}_{R}=({\psi_{0}}^{R},\psi_{1}^{R},\ldots,\psi_{n}^{R})$.
The mass matrix for the above class of models has the
following form
\begin{displaymath}
\mathbf{M}=
\left( \begin{array}{cccc}
\upsilon y_{0} & \upsilon y_{1} &  \ldots &  \upsilon y_{n} \\
0 & m_{1} &  \ldots &  0 \\
\vdots & 0 &  \ddots &  0 \\
0 & 0 &  \ldots &  m_{n} 
\end{array} \right)
\end{displaymath}


\section{Neutrinos in RS model}

For completeness of our analysis, we first briefly review the case of bulk
fermion spinors in the RS model \cite{Grossman:2000ra}.
This model consists of one positive (hidden) and one negative  tension
brane (where
the SM fields are confined)
placed on the fixed points ($y=0$, $L_{1}$) of a $S^{1}/Z_{2}$
orbifold (for details see \cite{Randall:1999ee}). In this case
the background geometry is described by eq.(1)  where  $\sigma(y)=k|y|$.
The convenient choice of variable, for the reasons
described in the previous section, is:

\be
z\equiv\frac{e^{ky}-1}{k} ~~~y\in[0,L_1]
\ee
Since in this model we have $(\sigma'(y))^2=k^2$ and $\sigma''(y)=
2kg(z) \left[\delta(z)-\delta(z-z_1)\right]$, the potential appearing
in eq.(8) of the Schr\"odinger equation that the wavefunction of the
right-handed bulk fermion is (for $z\ge 0$):
\be
{\rm}\hspace*{0.5cm} V_{R}(z)=\frac{\nu(\nu+1)k^{2}}{2[g(z)]^2}-
\frac{\nu}{2g(z)} 2k \left[\delta(z)-\delta(z-z_1)\right]
\ee
Here we have defined
$g(z)\equiv kz+1$ and $z_1\equiv z(L_1)$.

This potential always
gives rise to a (massless) zero mode.  It is given by
\be
\hat{f}_{0}^{R}(z)=\frac{A}{[g(z)]^{\nu}}
\ee

From the above expression it seems that the zero mode is always
localized on the positive tension brane for all values of
$\nu$. Nevertheless, by taking the second brane to infinity, we find that the zero
mode is normalizable  in the case that $\nu> \frac{1}{2}$ and that it
fails to be normalizable  when $0 \le \nu \leq \frac{1}{2}$.
The above, as we mentioned, shows that only when $\nu> \frac{1}{2}$
the zero mode is localized on the first brane. For $\nu= \frac{1}{2}$
there is no localization and for $ 0 \le \nu < \frac{1}{2}$ it is localized
on the negative brane. Another way to see the above is to find  the coupling of the KK states to mater
of a ``test'' brane as a function of the distance from the first
(hidden) brane. From eq.(13) we can find that in the case of $\nu > \frac{1}{2}$ the
coupling decreases as we go away from the first brane, on the other
hand it is constant when $\nu=\frac{1}{2}$ (no localization), and increases when
$0 \le \nu< \frac{1}{2}$ (localization on the second brane).
In any case as we previously mentioned the ``natural'' value for $\nu$
can be considered to be greater than unity (having already restricted
ourselves in the region $\nu > 0$) and thus we will assume in
the following discussions that the right-handed zero mode is always
localized on the hidden positive tension brane and we will  briefly discuss
the rest possibilities. Note that the all the following expressions for the
masses and the coupling are valid under the assumption that  $\nu >
\frac{1}{2}$, as the  results for the rest of the parameter space are different.
In this case  the normalization constant is $A\simeq \sqrt{k
(\nu-\frac{1}{2})}$.

Apart from the zero mode we have to consider the left and right-handed
KK modes which correspond to solutions for $m_{n}>0$. The solutions
for the right-handed wavefunctions in
this case are given in terms of Bessel
functions \footnote{Note that in the case that $\nu=N+\frac{1}{2}$, where $N$ is an
integer, the two linearly independent solutions are: $J_{N+1}$ and
$Y_{N+1}$. Although for our calculations we have assumed that
$ \nu \neq N +\frac{1}{2}$, all the results for the mass spectrum and the
couplings are valid also in the special cases when  $\nu=N+\frac{1}{2}$.}:

\be
{\hat{f}}^{R}_{n}(z)=\sqrt{\frac{g(z)}{k}}\left[A J_{\nu+\frac{1}{2}}\left(\frac{m_n}{k}g(z)\right)+B
J_{- \nu -\frac{1}{2}}\left(\frac{m_n}{k}g(z)\right)\right]
\ee
These solutions must obey the following  boundary conditions:
\ba
{{\hat f}^{R}_{n}}~'({0}^{+})+\frac{k \nu}{g(0)} {\hat f }^{R}_{n}(0)=0 \nonumber \\
{{\hat f}^{R}_{n}}~'({z_{1}}^{-})+\frac{k \nu}{g(z_{1})} {\hat f }^{R}_{n}(z_{1})=0
\ea

The wave functions of the left-handed KK states can be easily
extracted from eq.(10).
The  boundary conditions give a 2~x~2 system for $A$,$B$ which, in order
to have a nontrivial solution, should  have vanishing determinant. This
gives the quantization of the spectrum.
For $\nu> \frac{1}{2}$ the quantization condition  can be approximated by a simpler one:
$J_{\nu-\frac{1}{2}}\left(\frac{m_{n}g(z_{1})}{k}\right)=0$
This implies that the KK spectrum of the bulk state is:

\be
m_{n}=\xi_{n}~k~e^{-k L_{1}}
\ee
(for $n\ge{1}$), where $\xi_{n}$ in the n-th root
of $ J_{\nu-\frac{1}{2}}(x)$ . This means that if one is interested in solving the
hierarchy in the context of this model, i.e. $w \equiv e^{-k L_{1}}\sim 10^{-15}$
the mass of the first  bulk spinor KK state will be of the order of
1 TeV and the spacing between the tower will be of the same order.
To, summarize the spectrum in this case consists of a chiral
right-handed massless zero mode and a tower of Dirac KK states with
masses that start from 1 TeV (if a solution of the hierarchy is required)
with $\sim$1 TeV spacing.
The other important point for the phenomenology is the coupling of the
bulk spinors to the SM neutrino.
It is easy, using eq.(13), to find that the zero mode couples as
\be
\upsilon ~ y_{0}=\upsilon ~ Y_{5} ~ \sqrt{k (\nu-\frac{1}{2})}~~ \left(
\frac{1}{g(z_{1})} \right)^{\nu-\frac{1}{2}} \simeq \upsilon ~ \sqrt{\nu-\frac{1}{2}}~~ w^{\nu-\frac{1}{2}}
\ee
since, the hierarchy factor is defined as $w \equiv
\frac{1}{g(z_{1})}$ and, as mentioned in the previous section,  $Y_{5} \sim
\frac{1}{\sqrt{k}}$, $\upsilon \sim 10^2$ GeV and $\nu >
\frac{1}{2}$. 

In a similar fashion we can find  the couplings of the SM neutrino to
bulk KK states. In this particular model it turns out that this
coupling does not depend on the fermion mass or the size of the orbifold  and thus
it is a constant. By a simple calculation we find  that 

\be
\upsilon y_{n}~\simeq \sqrt{2} \upsilon ~ Y_{5} \sqrt{k} \simeq
\sqrt{2} ~ \upsilon 
\ee

Thus from the above we  see that the KK tower couples to SM neutrino with a TeV
strength.
In order to find the mass eigenstates and the mixing between the SM
neutrino and the sterile bulk modes one has to diagonalize the matrix
$\mathbf{M M^{\dag}}$ (actually one finds the squares of the mass
eigenvalues). By performing the above diagonalization,
 choosing  $ e^{-k L_{1}}\sim 10^{-15}$  it turns out that the
mass of the neutrino will be of the order
$m_{\nu} \sim 10^{2} ~(10^{-15})^{\nu-\frac{1}{2}}$  (e.g. for $\nu=\frac{3}{2}$, $m_{\nu}
\sim 10^{-4}$ eV), and the masses of the bulk states are of the order of
1TeV with a 1TeV spacing. From the last calculations it appears that one
can easily create a small neutrino mass and at the same time arrange
for  the desired mass hierarchy when $\nu > \frac{1}{2}$. 
 Apart from creating  small masses, one has to  check that the mixing between the SM neutrino and
the KK tower is  small enough  so that there is no conflict with phenomenology. It was shown in
Ref.\cite{Grossman:2000ra} that this can be done for this model
without fine-tuning. Note that the parameter space: $\nu \le
\frac{1}{2}$ (including negative values) is not of interest in the
present discussion \footnote{This could be of particular interest if
one uses the above mechanism to localize SM fermions on the negative
tension brane and in the same time solving the hierarchy problem
(e.g. see Ref.\cite{delAguila:2000kb}).} since it would be impossible to solve the hierarchy
problem and in the same time to assign small masses to neutrinos.


\section{Neutrinos in $''++-''$ model}

Since we are interested in studying the characteristics of bulk
fermion modes in multi-brane configurations we add to the $''+-''$ RS
model another positive tension brane
 where now SM fields will be confined. Thus we  end up with two different
configurations: the $''++-''$ model which will be the subject of this
section
and the  $''+-+''$ model which will be the subject of the next section.

The $''++-''$ model consists of two positive  and one negative tension
brane. The first positive brane is placed on the origin of the orbifold
at $y=0$ the second (where the SM fields are confined), which is freely moving, is place at $y=L_{1}$ and
the negative brane is placed at the second fixed point of the orbifold
at $y=L_{2}$.

In the present model the convenient choice of  variables  is defined as:
\be
\renewcommand{\arraystretch}{1.5}
z\equiv\left\{\begin{array}{cl}\frac{2e^{k_{1}L_1}-1}
{k_{1}}&y\in[0,L_1]\\\frac{e^{k_{2}(y-L_1)+k_{1}L{1}}}{k_{2}}+\frac{e^{k_{1}L_{1}}-1}
{k_{1}}-\frac{e^{k_{1}L_1}}
{k_{2}}&y\in[L_1,L_2]\end{array}\right.
\
\ee
Note the presence of two bulk curvatures, namely $k_{1}$ and $k_{2}$ 
in this model, which is the price that we have to pay in order to place
two positive branes next to each other  ($k_{1}<k_{2}$ but with $k_{1}
\sim k_{2}$ so that we don't introduce another hierarchy. For details see Ref.\cite{Kogan:2000xc}).
In terms of the new variables we can find that the potential
$V_{R}(z)$ of the Schr\"ondiger equation that corresponds to the present
model has the form (for $z \ge 0$):

\ba
\hspace*{0.5cm} V_{R}(z)=&\frac{\nu (\nu+1)}{2[g(z)]^2}(k_{1}^2(\theta(z)-\theta(z-z_{1}))+k_{2}^2(\theta(z-z_{1})-\theta(z-z_{2})))\nonumber\\&-
\frac{\nu}{2g(z)} 2\left[k_{1}\delta(z)+\frac{(k_{2}-k_{1})}{2}\delta(z-z_1)-k_{2}\delta(z-z_2)\right]
\ea
since $\sigma''(y)=2 g(z) \left[k_{1}\delta(z)+\frac{(k_{2}-k_{1})}{2}\delta(z-z_1)-k_{2}\delta(z-z_2)\right]$
and  $(\sigma'(y))^{2}=k_{1}^2$ for $y\in[0,L_{1}]$ and 
 $(\sigma'(y))^{2}=k_{2}^2$ for $y\in[L_{1},L_{2}]$. 
The function $g(z)$ is defined as: 
\be
\renewcommand{\arraystretch}{1.5}
g(z)=\left\{\begin{array}{cl}{k_{1}z+1}&z\in[0,z_{1}]\\{k_{2}(z-z_{1})+k_{1}z_{1}+1}&z\in[z_{1},z_{2}]\end{array}\right.
\
\ee
where $z_{0}=0$, $z_1=z(L_1)$ and $z_{2}=z(L_{2})$ are the positions of the
branes in terms of the new variables.

This potential always
gives rise to a (massless) zero mode whose wavefunction  is given by
\be
\hat{f}_{0}^{R}(z)=\frac{A}{[g(z)]^{\nu}}
\ee

The discussion of the previous section  about the state localization applies
in this model as well . For $\nu > \frac{1}{2}$ the zero mode is
localized on the first brane. For the case  $\nu = \frac{1}{2}$
 there is no localization again. For  $\nu < \frac{1}{2}$ it is localized on
the negative tension brane, as expected. 
In the case  $\nu > \frac{1}{2}$  we find that the normalization
factor of the zero mode is
$A\simeq\sqrt{k_{1}(\nu-\frac{1}{2})}$ which is the same as in the
case of RS for $k=k_{1}$ (not surprisingly since it is strongly localized on the
first brane).   

The wavefunctions for the right-handed  KK modes are given in terms of Bessel
functions. For $y$ lying in the regions ${\bf A}\equiv\left[0,L_1\right]$ and
${\bf B}\equiv\left[L_1,L_2\right]$, we have:

\be
\hat{\Psi}^{(n)}\left\{\begin{array}{cc}{\bf A}\\{\bf 
B}\end{array}\right\}=\left\{\begin{array}{cc}\sqrt{\frac{g(z)}{k_{1}}}\left[A_{1}J_{\nu+\frac{1}{2}}\left(\frac{m_n}{k_{1}}g(z)\right)+B_{1}J_{-\nu-\frac{1}{2}}\left(\frac{m_n}{k_{1}}g(z)\right)\right]\\\sqrt{\frac{g(z)}{k_{2}}}\left[A_
{2}J_{\nu+\frac{1}{2}}\left(\frac{m_n}{k_{2}}g(z)\right)+B_{2}J_{-\nu-\frac{1}{2}}\left(\frac{m_n}{k_{2}}g(z)\right)\right]\end{array}\right\}
\ee

with boundary conditions:

\ba
{{\hat{f}}^{R}_{n}}~'({0}^{+})+\frac{k_{1} \nu}{g(0)}{\hat{f}}^{R}_{n}(0)=0
\nonumber \\
{{\hat f}^{R}_{n}}({z_{1}}^{+})-{\hat f}^{R}_{n}({z_{1}}^{-})=0 \nonumber \\
{{\hat f}^{R}_{n}}~'({z_{1}}^{+})-{{\hat f}^{R}_{n}}~'({z_{1}}^{-})-\frac{2 
\nu}{g(z_{1})}\left( \frac{k_{2}-k_{1}}{2} \right) {\hat f}^{R}_{n}(z_{1})=0 \nonumber \\
{{\hat f}^{R}_{n}}~'({z_{2}}^{-})+\frac{k_{2} \nu}{g(z_{2})}{\hat f}^{R}_{n}(z_{2})=0
\ea

The above boundary conditions result  to a 4~x~4 homogeneous system for
$A_{1}$, $B_{1}$, $A_{2}$ and $B_{2}$ which, in order to have a nontrivial
solution, should have a vanishing determinant. This imposes a quantization condition from
which we are able to extract the mass spectrum of the bulk spinor. The
spectrum consists, apart from the chiral (right-handed) zero mode
(massless) which was mentioned earlier, of
a tower of Dirac KK modes.

In this case in order to provide a solution to the hierarchy problem we
have to arrange the distance between the first two branes so that 
we create the desired hierarchy $w$. In the
present model we have an additional parameter which is the distance
between the second and the third brane $x\equiv k_{2}(L_{2}-L_{1})$
For the region where $x\gtsim 1$ we can find analytically that all the masses the KK tower (including the
first's) scale the same way as we vary the length of the orbifold
$L_{2}$ : 

\be
m_{n}=\zeta_{n}~wk_{2}~e^{-k_{2}L_{2}}
\ee
 where $\zeta_{n}$ is the n-th root of $J_{\nu-\frac{1}{2}}(x)=0$.

In the region $x< 1$ the previous relation for the mass
spectrum breaks down. This is expected since for $x=0$
($L_{2}=L_{1}$)  the $''++-''$ model becomes $''+-''$ (RS) and
the quantization condition becomes approximately
$J_{\nu-\frac{1}{2}}\left(\frac{m}{k_{1}}g(z_1)\right)=0$, which is
identical to the RS condition. So for
$0\leq x\leq 1$ the quantization condition (and thus the mass spectrum) interpolates between
the previous two relations.

Let us now turn to the coupling between the SM neutrino which lives on
the second positive brane with the bulk right-handed zero mode and the
rest of the KK tower. We can easily derive that zero mode couples in
that same way  as in the RS case (the normalization of the zero mode
is approximately the same):
\be
 \upsilon ~ y_{0}=\upsilon ~ Y_{5} ~ \sqrt{k_{1}(\nu-\frac{1}{2})}~~
\left(\frac{1}{(g(z_{1}))}\right)^{\nu-\frac{1}{2}} \simeq \upsilon ~  \sqrt{\nu-\frac{1}{2}}~~
w^{\nu-\frac{1}{2}}
\ee
where, as we previously mentioned,   $Y_{5} \sim
\frac{1}{\sqrt{k_{1}}}$,~ $\upsilon \sim 10^{2}$ GeV, $\nu >
\frac{1}{2}$ and $w \equiv \frac{1}{g(z_{1})}$.
On the other hand the coupling of the SM neutrino to bulk KK states is
given by:

\be
 \upsilon ~ y_{n}\sim \upsilon ~ \sqrt{\nu} 
\left(\frac{k_{2}}{k_{1}}\right)^{3/2} \frac{8\zeta_{n}^2}{J_{\nu+\frac{1}{2}}\left(\zeta_{n}\right)}~e^{-3x}
\ee
where $\zeta_{n}$ is the n-th root of $J_{\nu-\frac{1}{2}}(x)=0$.

All approximations become better away from  $\nu = \frac{1}{2}$, $x=0$,
and for higher KK levels.
Note the strong  suppression in the coupling scaling law. This rapid
decrease, which is distinct among the models that we will consider,
also appears in the coupling (to matter) behaviour of the graviton KK
states and  a detailed explanation can be found in Ref.\cite{Kogan:2000xc}.  

Thus from the above we conclude that for $\nu > \frac{1}{2}$ the
phenomenology of this model resembles, in the general characteristics,
the one of RS. Of course in the present model there is an extra
parameter, $x$, which controls the details of the masses and couplings
of the KK states. Since the zero mode coupling is independent of $x$
the general arguments of the previous section about creating small
neutrino masses apply here as well, at least for small $x$ . By
increasing $x$ we make the KK tower lighter, as we see from eq.(28), but
 we avoid large mixings between the SM neutrino and the left-handed
bulk states due to the fact that the coupling between the SM neutrino
and the right-handed bulk states drops much faster according with eq.(30).

Note that  in the case  $\nu < \frac{1}{2}$ (negative values included) the zero mode 
will be localized on the negative brane and thus one could arrange the
parameter $x$ so that the exponential suppression of the bulk
fermion's zero mode coupling  on the
second brane is such that gives small neutrino masses. Thus in this
case it seems that we are able to solve the hierarchy problem by
localizing the graviton on the first positive brane and in the same
time create small neutrino masses by localizing the bulk fermion zero
mode on the negative brane. Nevertheless, one should make sure that no
large mixings are induced in this case.


\section{Neutrinos in $''+-+''$ model}

We now turn to examine bulk spinors in the $''+-+''$ model, which was analyzed
in detail in Ref.\cite{Kogan:2000wc,Mouslopoulos:2000er}. The model consists of two positive tension
branes placed at the orbifold fixed points and a third, negative brane
which is freely moving in-between. SM field are considered to be
confined on the second positive brane.
Of course the
presence of a moving negative brane is problematic  since it gives
rise to a radion field with negative kinetic term (ghost state)
\cite{Dvali:2000km,Pilo:2000et} in the gravitational sector.
 Nevertheless  we are interested in the  general characteristics of
this model . The interesting feature of this
model is the bounce form of the warp factor which gives rise to an
ultralight graviton KK state as described in Ref.\cite{Kogan:2000wc}. It was shown in Ref.\cite{Kogan:2000vb}
that exactly this feature, of a bounce in the warp factor, can be
reproduced even in the absence of negative branes in the $''++''$
model where this is done by sacrificing the flatness of the branes
(the spacetime on the branes in this case is $AdS_{4}$). 
Thus we will handle the $''+-+''$ as a toy model since in this
case there can be simple analytical calculations of the coupling etc. The
general characteristics will persist in the $''++''$ case.

We are interested to see if this configuration as well as  an ultralight graviton supports an ultralight spinor
field. In order to see this we should check the form of the potential
of the Schr\"ondiger equation that the right-handed component obeys. 
We can easily find that the  potential is (for $z \ge 0$):
\be
{\rm }\hspace*{0.5cm} V_{R}(z)=\frac{\nu(\nu+1) k^2}{2[g(z)]^2}-
\frac{\nu}{2g(z)} 2k \left[\delta(z)+\delta(z-z_2)-\delta(z-z_1)\right]
\ee
since $(\sigma'(y))^2=k^2$ and $\sigma''(y)= 2k g(z) \left[\delta(z)+\delta(z-z_2)-\delta(z-z_1)\right]$.
The convenient choice of variables in this case is:
\be
\renewcommand{\arraystretch}{1.5}
z\equiv\left\{\begin{array}{cl}\frac{2e^{kL_1}-e^{2kL_1-ky}-
1}{k}&y\in[L_1,L_2]\\\frac{e^{ky}-1}{k}&y\in[0,L_1]\end{array}\right.
\
\ee
and the function $g(z)$ is defined as $
g(z)\equiv k\left\{z_1-\left||z|-z_1\right|\right\}+1$, where $z_1=z(L_1)$.

As in the previous cases,  the above  potential always supports a (massless) zero mode
with wavefunction of the form:

\be
\hat{f}_{0}^{R}(z)=\frac{A}{[g(z)]^{\nu}}
\ee

In this case the different localization behaviour as a function of
$\nu$ is the following: For $\nu > \frac{1}{2}$ the zero mode is
localized on the positive branes (thus fails
to be normalizable when we send the right positive brane to infinity
but is normalizable when we send both negative and positive to
infinity). For the case $\nu < \frac{1}{2}$  the localization of
the zero mode is on the negative tension brane, as expected. 
In the case  $\nu > \frac{1}{2}$  and for strong hierarchy $w$ we find that the normalization
factor of the zero mode is
$A\simeq\sqrt{k_{1}(\nu-\frac{1}{2})}$
~ (Note that in the case of ``weak'' hierarchy one should be careful
with the assumptions on which the approximations are based on
e.g. for $w=1$ the result must be divided $\sqrt{2}$).   
For the KK modes the solution is given in terms of Bessel
functions. For $y$ lying in the regions ${\bf A}\equiv\left[0,L_1\right]$ and
${\bf B}\equiv\left[L_1,L_2\right]$, we have:

\be
{\hat{f}}^{R}_{n}\left\{\begin{array}{cc}{\bf A}\\{\bf 
B}\end{array}\right\}=\sqrt{\frac{g(z)}{k}}\left[\left\{\begin{array}{cc}A_1\\B_
1\end{array}\right\}J_{\frac{1}{2}+\nu}\left(\frac{m_n}{k}g(z)\right)+\left\{\begin{array}{cc}A_
2\\B_2\end{array}\right\}J_{-\frac{1}{2}-\nu}\left(\frac{m_n}{k}g(z)\right)\right]
\ee

with boundary conditions:

\ba
{{{\hat{f}}^{R}}_{n}}~'({0}^{+})+\frac{k \nu}{g(0)}{{\hat{f}}^{R}}_{n}(0)=0
\nonumber \\
{{{\hat f}^{R}}_{n}}({z_{1}}^{+})-{{{\hat f}^{R}}_{n}}({z_{1}}^{-})=0 \nonumber \\
{{{\hat f}^{R}}_{n}}~'({z_{1}}^{+})-{{{\hat f}^{R}}_{n}}~'({z_{1}}^{-})-\frac{2 k
\nu}{g(z_{1})}{{\hat f}}^{R}_{n}(z_{1})=0 \nonumber \\
{{{\hat f}^{R}}_{n}}~'({z_{2}}^{-})-\frac{k \nu}{g(z_{2})}{\hat f}^{R}_{n}(z_{2})=0
\ea

The  boundary conditions give a 4~x~4 linear homogeneous system for
$A_{1}$, $B_{1}$, $A_{2}$ and $B_{2}$, which, in order to have a nontrivial
solution should have  vanishing determinant. This imposes a quantization condition from
which we are able to extract the mass spectrum of the bulk spinor. The
spectrum consists, apart from the chiral (right-handed) zero mode
(massless) which was mentioned earlier, by
a tower of Dirac KK modes. Nevertheless due to the fact that there are
two positive tension branes present in the model there are now two
``bound'' states in a similar fashion with
Ref.\cite{Kogan:2000wc,Kogan:2000vb} (for $\nu > \frac{1}{2}$). One is the the right-handed zero mode which is
massless and it is localized on the positive brane placed at the
origin of the orbifold and the second is the ultralight right-handed
first KK state which is localized on the second positive brane placed at
the other orbifold fixed point. This can be seen by examining the mass
spectrum and the coupling behaviour of the first KK state in
comparison with the rest of the tower.
  
Firstly let us examine the mass spectrum. In the case that we have
a hierarchy $w$ (where $w\equiv \frac{1}{g(z_{2})}=e^{-\sigma(L_{2})}$) we can find appropriate analytical expressions
for the mass spectrum.

For the first KK state
\be
m_1=\sqrt{4 {\nu}^2 -1 } ~kw~ e^{-(\nu+\frac{1}{2}) x}
\ee
and for the rest of the tower
\be
m_{n+1}= \xi_n ~ kw~ e^{-x} ~~~~~~n=1,2,3, \ldots
\ee
where $\xi_{2i+1}$ is the $(i+1)$-th root of $J_{\nu-\frac{1}{2}}(x)$ ($i=0,1,2,
\ldots$) and $\xi_{2i}$ is the $i$-th root of $J_{\nu+\frac{1}{2}}(x)$ ($i=1,2,3, \ldots$).
The above approximations become better away from the $\nu=\frac{1}{2}$
 , $x=0$ and for higher KK levels $n$.
The first mass is manifestly singled out from the rest of the KK tower
as it has an extra exponential suppression that depends on the mass of
the bulk fermion. By contrast the rest of the KK tower has only a
very small dependence on the mass of the bulk fermion thought the
root of the Bessel function $\xi_{n}=\xi_{n}(\nu)$ which turns out to
be just a linear dependence in $\nu$. Note there is a difference between the
graviton ultralight state (discussed in \cite{Kogan:2000wc,Mouslopoulos:2000er}) and this spinor state: In the case of
gravity the unltralight KK state the mass  scales as a function of $x$
was  $ e^{-2x}$, on the other hand the scaling law in the
case of the ultralight spinor is of the form  $ e^{-(\nu+\frac{1}{2})x}$.
 From the above it seems that the latter can be done much lighter that
the graviton first KK state for a given $x$ by increasing the parameter $\nu$. This
is easy to understand since the role of the mass term, with the kink
or multi-kink profile, is to localize the wavefunction
$\hat{f}(z)$. By increasing the parameter $\nu$ all we do is to force
the  absolute value of the wavefunction of the first KK state and the
massless right-handed zero
mode to become increasingly similar to each other: For example, in the symmetric configuration, the the difference between the
 zero mode and the first KK state wavefunctions comes from
the central region of the $''+-+''$ configuration, where the first KK state
wavefunction is zero (since it is antisymmetric) thought the
zero mode's is very small due to the exponential suppression of the wavefunction, but non zero.
 By increasing $\nu$ we force  the value of the zero mode wavefunction at the
middle point
to get closer to zero and thus to resemble even more the
first KK state,
something that appears in the mass spectrum as the  fact that the mass of
the first KK state is approaching to zero. On the other hand the mass
eigenvalues that correspond to the rest
of the tower of KK states will increase linearly their mass by increasing the $\nu$ parameter
since those are not bound states (the first mode has also such a linear
dependence in $\nu$ but it is negligible compared with the exponential
suppression associated with $\nu$ ).

Now let us turn to the behaviour of the coupling of the zero mode and
the KK states to matter living on the third (positive) brane.
As in the previous cases the right-handed zero mode couples to SM left-handed neutrino as
\be
\upsilon ~ y_{0}=\upsilon ~ Y_{5} ~ \sqrt{k (\nu-\frac{1}{2})}~~ {\left(
\frac{1}{g(z_{2})} \right) }^{\nu-\frac{1}{2}} \simeq \upsilon  ~ \sqrt{\nu-\frac{1}{2}}~~
{ w^{\nu-\frac{1}{2}}}
\ee
since $Y_{5} \sim \frac{1}{\sqrt{k}}$. 
From the above relationship we
see that the coupling of the zero mode to SM neutrino will 
generally be suppressed by the hierarchy factor to some power, the power depending on
the bulk fermion mass.  This way one may  readily obtain a very small
coupling.
 The coupling of the zero mode is independent of
$x$. This  is another way to see the localization of this mode
on the first brane (the normalization of the wavefunction is
effectively independent of $x$). Since this model supports a
second ``bound state'' (first KK state) which is localized on the
second brane, we expect something similar to occur in the coupling
behaviour of this state. Indeed, similarly to the graviton case \cite{Kogan:2000wc,Mouslopoulos:2000er}, we can
show that the coupling of this state to the SM neutrino for fixed $w$
is constant, i.e. independent of the $x$ parameter. Taking in account
the result of the graviton KK state $a_{1}=\frac{1}{wM_{Pl}}$ and by
comparing the graviton-matter and spinor matter coupling we can easily
see that the coupling of this special mode will  be of the order of
the electroweak scale:

\be
\upsilon ~y_{1}~\simeq \sqrt{\nu-\frac{1}{2}} ~ \upsilon 
\ee

Let us now consider the coupling of the rest right-handed KK states to
the SM neutrino. We  find that

\be
\upsilon ~y_{n}~\simeq \sqrt{\nu-\frac{1}{2}} ~ \upsilon ~ e^{-x} 
\ee
for $n=0,1,2...$.
From the above relationship we see that the rest of KK states will
generally have exponentially suppressed coupling compared to the first
special state.

The appearance of this special first ultralight and generally strongly
coupled KK state, as in the graviton
case, is going to have radical implication to  the phenomenology of
the model. Let us  consider the
following example suppose that $\nu=\frac{3}{2}$ and that we also require a hierarchy of
the order: $w \sim 10^{-15}$. In this case
the zero mode's coupling is  $ \upsilon y_{0} \sim  \upsilon
10^{-15} \simeq 10^{-4}$ eV a result independent of the $x$
parameter. On the other hand one can check that the rest of KK tower
will have masses $m_{n}\simeq 10^{3}~e^{-x}$ GeV (for $n=2,3...$) with coupling
$ \upsilon y_{n} \sim  \upsilon e^{-x}$. Up to this point the
phenomenology associated with this model is similar to the  RS case
i.e. tiny coupling of the right-handed and generally heavy KK states
with relatively strong coupling. However, taking in account the
special KK state, we have the possibility of obtaining  a much lighter state
with large coupling (effectively independent of how light this state
is). Having a light sterile state whose right-handed mode has strong
coupling to the SM neutrino is potentially dangerous. In such a case
we find that the dominant
contribution to the mass eigenstate  of the lightest mode
(neutrino) $\nu_{phys}$ will come from the left-handed component of
this special sterile mode and not the weak eigenstate $\nu$.
Of course something like this is not
acceptable since there are strict constrains for the mixing of SM
neutrino to sterile states. Since the mass spectrum depends
exponentially on the the $x$ parameter which determines the distance
between the branes, the above argument impose strong constraints on
it's  possible values. 

Finally, in the case that $\nu < \frac{1}{2}$ the bulk right-handed zero
mode is localized on the negative tension brane. In this case a new
possibility arises: By localizing the graviton wavefunction on the
first brane we can explain the SM gauge hierarchy (by setting $w$ to
the desired value ) and by localizing the bulk fermion zero mode on
the negative tension brane to induce small neutrino mass (for
appropriate value of the $x$ parameter) for the SM
neutrino which is confined on the right positive tension brane.  
 Note that for $\nu < \frac{1}{2}$ there is no special bulk spinor KK state
and thus there is no immediate danger of inducing large
neutrino mixing from such a state. However, the presence of the
ultralight graviton KK state is restricting our parameter space as following: 
 In order to solve the gauge hierarchy problem (assuming the SM on the
third brane) we have to fix the one
parameter of the model: $w\sim 10^{-15}$. Since the fermion zero mode is localized
on the intermediate negative brane we have to arrange the distance
between this and the third brane (for given $\nu$), $x$, so that the 
coupling is sufficiently  suppressed in order to give reasonable
neutrino masses. This implies that 
$e^{- |\nu-\frac{1}{2}| x} \sim 10^{-13}$,   if one considers the mass of
the neutrino of the order of $10^{-1}$ eV. From the bounds derived in \cite{Mouslopoulos:2000er}
 we find, for $k\sim10^{17}$ GeV, that in order the ultralight
graviton KK state not to induce modifications of gravity at distances
where Cavendish experiments take place and not to give visible
resonances to  $e^{+}e^{-}\rightarrow \mu^{+} \mu^{-}$ processes, we should have
$4.5<x<15$ or $x<1$. The latter implies certain restrictions to the
values of $\nu$: $1.5<-\nu<6.2$ or $-\nu>29.4$. In the above regions
it is possible to simultaneously create the gauge hierarchy and small
neutrino masses consistently, with the mechanism described earlier.


\section{Neutrinos in $''++''$ model}

As we mentioned in the previous section the $''++''$ model mimics the
interesting characteristics of the $''+-+''$ model without having any
negative tension brane. Thus since the warp factor has a bounce form
this model also supports an ultralight graviton as  was shown in
Ref.\cite{Kogan:2000vb,Karch:2000ct,Miemiec:2000eq,Schwartz:2000ip}. According to the previous discussion, we should expect
that the model will support a ultralight sterile neutrino as well. 
This can be easily shown again by considering the form of the
potential of the differential equation that the right-handed component
is obeying, which will again turn out to be
of the same form as the graviton. 
 
Now it turns out that for the construction of  such a $''++''$
configuration, it is essential to have $AdS_{4}$ geometry on both
branes (for details see Ref.\cite{Kogan:2000vb}). Thus in this case the background geometry is described by:

\be
ds^2=\frac{e^{-2\sigma(y)}}{(1-\frac{H^{2}x^{2}}{4})^2}\eta_{\mu\nu}dx^{\mu}dx^{\nu}
- dy^2
\ee
where the corresponding inverse vielbein is given by 
\be
{E}_{\alpha}^{A}=diag(e^{\sigma(y)}{(1-\frac{H^{2}x^{2}}{4})},e^{\sigma(y)}{(1-\frac{H^{2}x^{2}}{4})},e^{\sigma(y)}{(1-\frac{H^{2}x^{2}}{4})},e^{\sigma(y)}{(1-\frac{H^{2}x^{2}}{4})},1).
\ee

Since now the brane is no longer flat the previous calculations for
the action will be slightly modified. We briefly discuss these modifications.
Following the same steps of the flat case, we write $\Psi=\Psi_{R}+\Psi_{L}$ where
$\Psi_{R,L}=\frac{1}{2}(1\pm \gamma_{5})\Psi$. Since the connection part of the Lagrangian again doesn't give any
contribution (since the vielbein is again symmetric) the  action becomes:

\ba
S=\int \sqrt{\hat{G}} ~d^4x ~ dy  
\{  e^{-3\sigma} {{\hat{E}}}^{A}_{a}\left( \bar{\Psi}_{L}i\gamma^{a}\partial_{A} \Psi_{L} +
\bar{\Psi}_{R}i \gamma^{a} \partial_{A} \Psi_{R} \right) -  e^{-4\sigma} m
\frac{\sigma'(y)}{k} \left( \bar{\Psi}_{L}\Psi_{R} +
\bar{\Psi}_{R}\Psi_{L} \right) \nonumber \\ -\frac{1}{2}\left[ \bar{\Psi}_{L} (e^{-4\sigma}\partial_{y}+\partial_{y}e^{-4\sigma} ) \Psi_{R} -
\bar{\Psi}_{R}(e^{-4\sigma}\partial_{y}+\partial_{y}e^{-4\sigma} ) \Psi_{L}  \right] \}
\ea
where
$\hat{E}_{\alpha}^{A}={(1-\frac{H^{2}x^{2}}{4})}\delta_{\alpha}^{A}$
with ($a$,$A$=0,1,2,3) is the induced vielbein and $\hat{G}$ the
determinant of the induced metric.
For convenience and in order to be able to use results
of Ref.\cite{Kogan:2000vb}, we set $A(y)\equiv e^{-\sigma(y)}$ where

\be
A(y)=\frac{\cosh(k(y_{0}-|y|))}{\cosh(ky_{0})}
\ee
is the equivalent ``warp'' factor in this case, which is found by
considering the configuration gravitationally.
From the above relation it is clear that the ``warp'' factor has the
desired bounce form, with a minimum at $y_{0}$. 
The position of the minimum, something that it is going to be important
for the phenomenology of the model, is  defined from the relationship
$\tanh(ky_{0})\equiv\frac{k V_{1}}{|\Lambda|}$, where $V_{1}$ is the
tension of the first brane and $\Lambda$ is the five dimensional
cosmological constant. As we mentioned in the
introduction, the profile of the mass term, $\frac{A'(y)}{A(y)}$, in
the present model is not a simple combination of
$\theta$-functions. In particular there are two $\theta$-function
profiles near the orbifold fixed points  which give rise to the
positive branes, but there is also an intermediate kink profile of the
form $-\tanh(k(y-y_{0}))$ which is associated with the presence of the bounce
(this could give rise to a $''-''$ brane as a limit , resulting to the familiar $''+-+''$
configuration). Note that even though there is no brane at the position
of the minimum of the ``warp'' factor the kink profile is expected to
act in the same way, and thus induce localization of the fermion zero
mode in specific regions of the parameter space, exactly as in the $''+-+''$ model.

As in the flat-brane case, we can decompose the left-handed and
right-handed fermion fields into KK states with nontrivial profile
wavefunctions $f^{L}_{n},f^{R}_{n}$ (in respect to the fifth dimension)  in order to be able to bring the
Lagrangian into the form 

\be
S=\sum_{n} \int d^4x \sqrt{-\hat{g}}
\{\hat{E}_{\alpha}^{A}\bar{\psi}_{n}(x) i \gamma^{\alpha} \partial_{A} \psi_{n}(x) -
m_{n}\bar{\psi}_{n}(x) \psi_{n}(x) \}
\ee

where
the wavefunctions $f^{L}_{n}(y)$, $f^{R}_{n}(y)$  should obey the following equations

\ba
\left( -\partial_{y}
+\frac{m}{k}\frac{A'(y)}{A(y)}\right)f^{L}_{n}(y)=m_{n} A^{-1}(y) f^{R}_{n}(y)
\nonumber \\
\left( \partial_{y} +\frac{m}{k}\frac{A'(y)}{A(y)}\right)f^{R}_{n}(y)=m_{n}A^{-1}(y)f^{L}_{n}(y)
\ea

with the following orthogonality relations:

\be
\int_{-L}^{L} dy A^{-1}(y) {f^{L}_{m}}^{*}(y) f^{L}_{n}(y)=\int_{-L}^{L} dy A^{-1}(y) {f^{R}_{m}}^{*}(y) f^{R}_{n}(y)=\delta_{mn}
\ee

Again we solve the above system of differential equations by finding
the second order differential equation that it implies for the
right-handed component of the spinor.
It is always possible to make the coordinate transformation from y
coordinates to z coordinates  related through:
$\frac{dz}{dy}=A^{-1}(y)$
and bring the differential equations in the familiar form:

\be
\left\{-
\frac{1}{2}{\partial_z}^2+V_{R}(z)\right\}{\hat{f}^{R}}_{n}(z)=\frac{m_n^2}{2}
{\hat{f}^{R}}_{n}(z)
\ee
where we have defined $\hat{f}^{R}_{n}(z)=f^{R}_{n}(y)$
\ba
{\rm with}\hspace*{0.5cm} V_{R}(z)&=&\frac{\nu}{2}A(y)A''(y)+
\frac{{\nu}^{2}}{2}(A'(y))^{2} \nonumber \\
&=&-\frac{{\nu}^2\tilde{k}^{2}}{2}~+~\frac{\nu(\nu+1)\tilde{k}^2}{2}\frac{1}{\cos^{2}\left(\tilde{k}(|z|-z_{0})\right)}\cr      &-&k\nu\left[ \tanh(ky_{0})\delta(z)+\frac{\sinh(k(L-y_{0}))\cosh(k(L-y_{0}))}{\cosh^{2}(ky_{0})}
\delta(z-z_{1})\right] 
\ea
with $\tilde{k}$  defined as
$\tilde{k}\equiv{\frac{k}{\cosh(ky_{0})}}$.
The new variable $z$ is  related to the old one $y$ through the relationship:
\be
z\equiv {\rm sgn}(y)\frac{2}{\tilde{k}}\left[\arctan\left(\tanh(\frac{k(|y|-y_{0})}{2})\right)+\arctan\left(\tanh(\frac{ky_{0}}{2})\right)\right]
\
\ee
Thus in terms of the new coordinates, the branes are  placed at $z_{1}=0$
 and $z_{L}$, with the minimum of the potential  at $z_{0}={2 \over \tilde{k}}\arctan\left(\tanh(\frac{ky_{0}}{2})\right)$. Also note
that with this transformation the point $y=\infty$ is mapped to the
finite point $z_{\infty}={2 \over \tilde{k}}\left[{\pi \over 4} +
 \arctan\left(\tanh(\frac{ky_{0}}{2})\right)\right]$.

We can now proceed to the solution of the above equations for the
right-handed components, while the left-handed wavefunctions can be
easily evaluated using eq.(10) (taking in account the definition $A
\equiv e^{- \sigma(y)}$ and the change of variables). The zero
mode wavefunction is given by:
\be
\hat{f}_{0}^{R}(z)=\frac{C}{[\cos(\tilde{k}(z_{0}-|z|))]^{\nu}}
\ee
where $C$ is the normalization factor. If we send one of the two
branes to infinity (i.e. $z_{1} \rightarrow z_{\infty}$) and at the
same time keep $z_{0}$ fixed we find that the zero mode is
normalizable only in the cases where $\nu < \frac{1}{2}$. In the other
cases ($\nu \ge \frac{1}{2}$) the wavefunction fails to be
normalizable due to the fact that
it is too singular at $z_{\infty}$. Note though that in the case where
$\nu < \frac{1}{2}$ the first KK will not be special (i.e. will have
almost the same behaviour as the rest of the KK tower).
We also note that again  the zero mode is chiral i.e. there is no solution for
$f_{n}^{L}$ when $m_{0}=0$ that can satisfy the boundary conditions
(antisymmetric wavefunction). From the above we see that the
localization behaviour of zero mode in the $''++''$ model is the same 
as in the $''+-+''$ model. Note that for $\nu < \frac{1}{2}$ the zero
mode will be localized near $y_{0}$  despite the absence of any brane
at that point. This is because, as we mentioned, the $-
\frac{A'(y)}{A(y)}$ factor, which can be considered as the vacuum
expectation value of a scalar field, has a kink profile in the
neighbourhood of $y_{0}$ which induces the localization (this kink
becomes the negative tension brane in the flat brane limit). 
By considering cases with $m_{n}\neq0$, we find the wavefunctions  for the KK tower :

\be
\renewcommand{\arraystretch}{1.5}
\begin{array}{c}{\hat{f}^{R}}_{n}(z)=\cos^{\nu+1}(\tilde{k}(|z|-z_{0}))\left[C_{1}~F(\tilde{a}_{n},\tilde{b}_{n},\frac{1}{2};\sin^{2}(\tilde{k}(|z|-z_{0})))~~~~~~~~\right.
\\ \left.  ~~~~~~~~~~~~~~~~~~~+C_{2}~|\sin(\tilde{k}(|z|-z_{0}))|~F(\tilde{a}_{n}+\frac{1}{2},\tilde{b}_{n}+\frac{1}{2},\frac{3}{2};\sin^{2}(\tilde{k}(|z|-z_{0})))\right]
\end{array}
\ee
where
\ba
\tilde{a}_{n}=\frac{\nu+1}{2}+\frac{1}{2}\sqrt{\left(\frac{m_{n}}{\tilde{k}}\right)^2+{\nu}^2}
\cr
\tilde{b}_{n}=\frac{\nu+1}{2}-\frac{1}{2}\sqrt{\left(\frac{m_{n}}{\tilde{k}}\right)^2+{\nu}^2
}
\ea
The  boundary conditions are given by:
\ba
{\hat{f}^{R}}_{n}~'({0}^{+})+k \nu \tanh(k y_{0}){\hat{f}^{R}}_{n}(0)=0
\nonumber \\
{\hat{f}}^{R}_{n}~'({z_{L}}^{-})-k \nu \frac{\sinh(k(L-y_{0}))}{\cosh(k y_{0})}{\hat{f}^{R}}_{n}(z_{L})=0
\ea
the above conditions determine the mass spectrum of the KK states. 
By studying the mass spectrum of the KK states it turns out that
it  has a special first
mode similar to the one of the $''+-+''$ model as expected. 
For example, for the symmetric configuration ($w=1$), by approximation we can analytically find the following expressions
for the mass of this special state :

\be
m_1=2\sqrt{4 {\nu}^2+3}~k~\left(e^{-ky_{0}}\right)^{\nu+\frac{1}{2}}
\ee
In contrast, the masses of the next levels are given by the formulae:
For odd states
\be
m_{n}=2  \sqrt{(n+1)(n+1+\nu)}~k ~e^{-ky_{0}}
\ee
with $n=0,1,2,...$, and for even states
\be
m_{n}= 2 \sqrt{( n + \frac{3}{2} )( n + \frac{3}{2}+ \nu)} ~k~ e^{-ky_{0}}
\ee
with $n=0,1,2,...$.

Again the  mass of the first KK state is manifestly singled out from the rest of the KK tower
as it has an extra exponential suppression that depends on the mass of
the bulk fermion. The above characteristics persist in the more
physically interesting  asymmetric
case ($w<<1$). In this case we find $m_{1} \sim 
m_{1}^{0}(\nu)~kw~(e^{-kx})^{\nu+\frac{1}{2}}$ and $m_{n} \sim m_{n}^{0}(\nu)~  kw~ e^{-kx}$ where
$x=L_{1}-y_{0}$ is the distance of the second brane from the minimum
of the warp factor, and $m_{1}^{0} \sim 1$, $m_{n}^{0}$ is linear to
$n$ (for large values)  and has a $\sim \sqrt{\nu}$ dependence. 
  The first KK state is
localized on the second positive brane and, as in the case of
$''+-+''$  model, its coupling to SM neutrinos
remains constant if we keep the hierarchy parameter $w$ fixed.
The phenomenology of this model will be similar to the
$''+-+''$ model and thus we do not consider it
separately.


\section{ Bigravity and Bulk spinors }

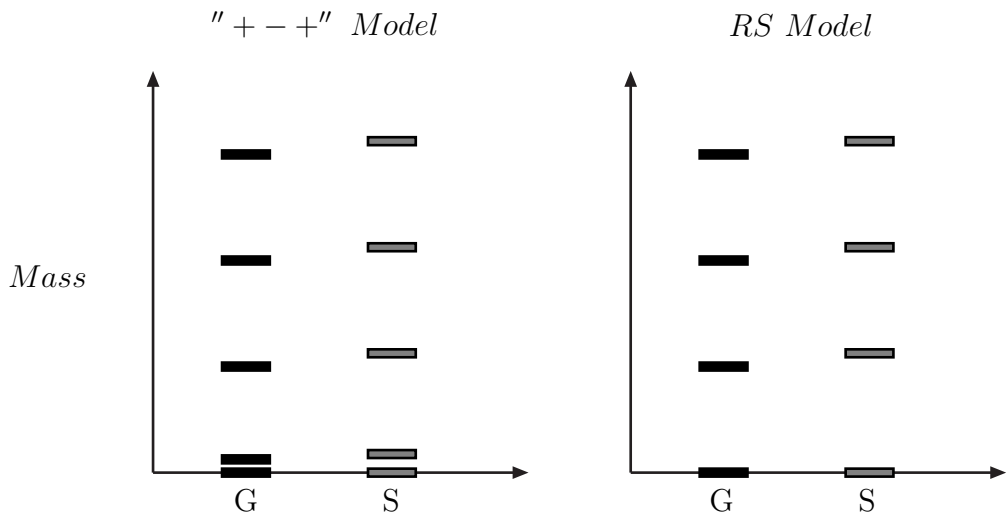
\begin{figure}[t]
\begin{center}
\begin{picture}(300,200)(0,70)

\SetWidth{1}
\LongArrow(-20,100)(-20,250)
\LongArrow(160,100)(160,250)

\SetWidth{1}
\LongArrow(-20,100)(120,100)
\LongArrow(160,100)(300,100)

\Text(-60,175)[c]{$Mass$}
\Text(45,270)[c]{$''+-+''~Model$}
\Text(225,270)[c]{$RS~Model$}
\Text(15,90)[c]{G}
\Text(70,90)[c]{S}
\Text(195,90)[c]{G}
\Text(250,90)[c]{S}



\GBoxc(15,220)(18,3){0}
\GBoxc(15,180)(18,3){0}
\GBoxc(15,140)(18,3){0}

\GBoxc(15,105)(18,3){0}

\GBoxc(15,100)(18,3){0}




\GBoxc(70,225)(18,3){0.5}
\GBoxc(70,185)(18,3){0.5}
\GBoxc(70,145)(18,3){0.5}

\GBoxc(70,107)(18,3){0.5}

\GBoxc(70,100)(18,3){0.5}




\GBoxc(195,220)(18,3){0}
\GBoxc(195,180)(18,3){0}
\GBoxc(195,140)(18,3){0}

\GBoxc(195,100)(18,3){0}




\GBoxc(250,225)(18,3){0.5}
\GBoxc(250,185)(18,3){0.5}
\GBoxc(250,145)(18,3){0.5}

\GBoxc(250,100)(18,3){0.5}


\end{picture}
\end{center}

\caption{On the left, the mass spectrum of the graviton (G) (first column)
and bulk spinor (S) (second column)
KK states in the $''+-+''$-Bigravity model. On the right, for comparison, the
corresponding spectrum for the case of RS model. The figures are not in
scale and the details of the bulk spinor mass spectrum depend on the
additional parameter $\nu$ and thus in the above figure we have
assumed $\nu \simeq \frac{3}{2}$ (for higher value of $\nu$ we could have the fermion
first KK state to be lighter that the first graviton KK state).}

\end{figure}

 Bigravity \cite{Kogan:2000wc,Mouslopoulos:2000er,Kogan:2000vb}
(multigravity \cite{Gregory:2000jc,Kogan:2000xc}) is the possibility
that gravitational interactions do not exclusively come from a massless graviton
, but instead they can be the net effect of a massless graviton
and one or more KK states (continuum of KK states) or even a single  massive
KK mode \cite{Kogan:2000vb,Karch:2000ct} without conflict with General
Relativity predictions \cite{Kogan:2000uy,Porrati:2001cp}.   
This is based on the different
scaling laws of the mass between the first and the rest of KK states. 
Since the first graviton KK state has mass with an additional exponential
suppression we can realize
the scenario that the first KK state is so light that its wavelength
is of the order of the observable universe and thus any observable
effect of its non-vanishing mass to be out of the experimental reach, and
in the same time the rest of the KK tower has masses above the scale
that Cavendish experiments have tested Newtonian gravity at small distances.

In the two previous sections we have shown that in the case of models,
where Bigravity can be realized, there is also an ultralight KK that
corresponds to the bulk spinor assuming the existence of the mass term for the
bulk spinor  that appears in eq.(2) (with $\nu > \frac{1}{2}$) . The subject of this section is to
investigate if the two above possibilities are compatible: Can we have
a Bigravity scenario and a consistent neutrino phenomenology? For the
sake of simplicity the discussion below will be concentrated to the
$''+-+''$ model but, analogous arguments
should apply to the case of $''++''$ model. 

Let us review briefly the Bigravity scenario.
The graviton ultralight first KK state has mass
$m^{(G)}_{1}=2wk~e^{-2x}$ and coupling $a^{(G)}_{1}=\frac{1}{wM_{Pl}}$. The rest
of the KK tower has masses :
\be
m_{n+1}^{(G)}= \xi_n ~ kw~ e^{-x} ~~~~~~n=1,2,3, \ldots
\ee
where $\xi_{2i+1}$ is the $(i+1)$-th root of $J_{1}(x)$ ($i=0,1,2,
\ldots$) and $\xi_{2i}$ is the $i$-th root of $J_{2}(x)$ ($i=1,2,3,
\ldots$). The couplings of these states scale as $a_{n}
\propto e^{-x}$.
In order to achieve the Bigravity scenario, as we mentioned in the
beginning of this section, we have the following  constrains on the
range of masses of the KK states: $m^{(G)}_{1}<10^{-31}{\rm eV}$
or $m^{(G)}_{2}>10^{-4}{\rm eV}$ (where Planck suppression is considered for
the ``continuum '' of states above $10^{-4}$eV). Our exotic scheme corresponds 
to the choice $m^{(G)}_1\approx 10^{-31}{\rm eV}$ and $m^{(G)}_2>10^{-4}{\rm eV}$. In this case, for 
length scales less than $10^{26}{\rm cm}$ gravity is generated by the exchange of {\it 
both} the massless graviton and the first KK mode. This implies, (taking into 
account the different coupling
suppressions of the massless graviton and the first KK state) that
the gravitational coupling as we measure it is related to the
parameters of our model by:
\be
\frac{1}{M_{\rm
Pl}^2}=\frac{1}{M_{5}^2}\left(1+\frac{1}{w^2}\right)\approx
\frac{1}{(w M_{5})^2} \Rightarrow M_{\rm Pl}\approx w M_{5}
\ee

We see that the mass scale on our brane, $w M_{5}$, is now the Planck scale so, 
although the ``warp'' factor, $w$, may still be small (i.e. the fundamental 
scale  $M_{5} >>M_{Planck}$), we do not now solve  the Planck hierarchy
problem. Using the equations for the mass spectrum  and assuming as before that $k\approx M_{5}$, 
we find that $m_{1}=2~kw~e^{-2x}\approx M_{Planck}e^{-2x}$. For $m_1=10^{-31}{\rm eV}$ we 
have $m_{2}\approx 10^{-2}{\rm eV}$. This comfortably satisfies the
bound $m>10^{-4}{\rm eV}$. 

Now let us see what this implies for the neutrino physics.
Let us first consider the case where $\nu > \frac{1}{2}$.
In this case, as mentioned above, there will be also an ultralight bulk
fermion KK mode. By forcing the graviton first mode to have  a
tiny mass we also force the first spinor KK state to become very light
 (even in the best case where $\nu \rightarrow \frac{1}{2}$ this state will
have mass of the order of a fraction of eV. For larger $\nu$ becomes
even lighter) since they are related through: $m_{1} \sim
e^{(\frac{3}{2}-\nu)x}~ m^{G}_{1}$. This  is unacceptable since this mode has a constant coupling of the
order of the weak scale, $\upsilon$, which will induce large mixing of
the neutrino with the left-handed component of this state.
However note that in the limit $\nu \rightarrow \frac{1}{2}$ the
special first fermion KK mode will become a normal one losing it's localization
and thus the above may not apply. Unfortunately, this is not the case as
in this limit  the fermion
zero mode is delocalized (for $\nu = \frac{1}{2}$ the coupling is
constant across the extra dimension) giving no possibility of inducing
small neutrino masses (since the compactification volume is very
small)\footnote{Trying to use this window of the  parameter space 
seems like a fine-tuning though since
one needs to delocalize the fermions first KK state enough in order to
have small coupling to SM neutrino and on the other hand to prevent
the delocalization of the fermion zero mode with the same mechanism
($\nu \rightarrow \frac{1}{2}$).}.
 
 Despite the  severe constraints in the above scenario, due to the
presence of the ultralight bulk fermion KK state,
by no means the Bigravity scenario is excluded in the the case of
$\nu>\frac{1}{2}$ since one can always consider the possibility of
placing the SM on  a  brane (with tiny tension so that
the background is not altered) between the negative and the second
positive brane, so that the coupling of the bulk fermion first KK
state to SM neutrinos  is sufficiently small while  a part of
gravitational interactions will still be generated from the ultralight
graviton first KK state.

Let us turn now to the case $0 \le \nu < \frac{1}{2}$. In this case there is
no special bulk spinor KK state and thus the above arguments do not
apply. In this case  the bulk fermion zero mode is
localized on the negative tension brane. Nevertheless, in order for the
Bigravity scenario to be possible we should have $x \simeq 60$. If we
now try to generate neutrino masses with the  mechanism described in
the first section we will find that the coupling of the zero mode to
SM neutrino will be by far too small to provide  consistent results.
Thus in this case although Bigravity is realized we cannot generate
neutrino masses consistently. 

The case of  $\nu=\frac{1}{2}$ is of no interest since in this case
there is no localization (the fermion zero mode has constant coupling
across the extra dimension). The case where $\nu < 0$ resembles the $0
\le \nu < \frac{1}{2}$ case, with the only difference that the
localization on negative branes will be even sharpner, making the
situation even worse.


\section{Discussion and conclusions}

We have studied bulk fermion fields in various multi-brane  models with localized
gravity. The chiral zero model that these models support can be
identified as a right-handed sterile neutrino. In this case small
neutrino Dirac masses can naturally appear due
to an analogous (to graviton) localization of the bulk fermion zero
mode wavefunction without
invoking a see-saw mechanism.  For models in which the localization
of the fermion zero mode is induced by the same scalar field that
forms the
branes  the localization behaviour of this mode can resemble
the graviton's at least in a region of  parameter space.
The latter implies that the $''+-+''$, $''++''$ models can support,
in addition to  the ultralight graviton KK state, an
ultralight localized and strongly coupled  bulk fermion KK mode.
This fermion state, when exists, imposes even more severe
constrains on the parameter space of $''+-+''$ and $''++''$
models. In the case that one requires the Bigravity be realized
the light fermion KK mode can induce too large mixing between the neutrino
and the KK tower and thus it restricts even more the allowed parameter
space of the relevant models.
 
As a general remark  we see that the appearance of
Multi-localization \cite{moriond} in the Multli-Brane world
picture  and its relation to the existence of ultralight states in
the KK spectrum is not a characteristic of the graviton only, but can
also occur in spin $0$, $\frac{1}{2}$ and $\frac{3}{2}$ bulk states with
appropriate bulk mass terms \cite{next}. Vector
fields on the other hand can have also ultralight states but in this case a
coupling to a bulk scalar field is necessary  in order to
achieve localization \cite{Kehagias:2000au}. 
 
The appearance of ultralight KK states with the additional
characteristic, in the case of spin $\frac{1}{2}$ fields, where the
left-handed and the right-handed components are localized in
different places in the extra dimension, can give new interesting
possibilities which will be discussed in another publication \cite{next}.

\vspace{0.7cm}


\textbf{Acknowledgments:}
I would like to thank Ian I. Kogan   
for useful discussions and comments, Antonios Papazoglou for careful
reading of the manuscript - useful discussions and comments.
I am grateful to Graham G. Ross for suggesting
this work, important discussions and careful reading of the
manuscript. I would also like to thank Shinsuke Kawai for useful discussions.  
This work is supported by the Greek State
Scholarship Foundation (IKY) scholarship \mbox{No. 
8117781027}.


\end{document}